\documentclass{article}
\usepackage{spconfa4,amsmath,graphicx}
\usepackage{amssymb}
\usepackage{xcolor}
\usepackage{algorithm}
\usepackage{algpseudocode}
\usepackage{svg}
\usepackage{tabularray}
\usepackage{booktabs}
\usepackage{multirow}
\newcommand{\myparagraph}[1]{\noindent\textbf{#1} \hspace{.12cm}} 

\title{Test-time adaptation for speech enhancement with an autoregressive speech prior}

\name{Sofiene Kammoun$^{1}$  \qquad Simon Leglaive$^{1}$ \qquad Xavier Alameda-Pineda$^{2}$ \qquad Timo Gerkmann$^3$}
  \address{$^{1}$CentraleSupélec, IETR (UMR CNRS 6164), France \\
      $^{2}$Inria at Univ. Grenoble Alpes, CNRS, LJK, France \\
      $^{3}$Signal Processing Group, University of Hamburg, Germany} 

\begin{document}
\ninept
\maketitle
\begin{abstract}
Test-time adaptation (TTA) offers a promising direction for improving speech enhancement models under mismatched acoustic conditions, without requiring access to labeled target data. In this work, we propose a single-utterance TTA method that regularizes a pretrained speech enhancement model using an autoregressive prior trained on clean speech latent representations extracted from a neural audio codec. Adaptation is performed by minimizing the Kullback-Leibler divergence between the enhanced speech distribution and the clean speech prior. Experiments across multiple noisy speech datasets show consistent improvements in speech quality, particularly under training-testing noise mismatch conditions. Code and audio examples are available online.\footnote{https://sofienekammoun.github.io/TAAP-SE/\\ This research was partly supported by the ANR project ANR-23-CE23-0009.}
\end{abstract}

\section{Introduction}

Speech enhancement (SE) aims to recover a clean speech signal from a degraded recording, typically affected by environmental noise. Recent SE systems predominantly rely on supervised deep learning models trained on paired noisy-clean speech data. While these approaches achieve strong performance under matched training and testing conditions, they can degrade significantly when the acoustic characteristics encountered at test time differ greatly from those seen during training, or when the input signal contains unseen corruption types such as reverberation, bandwidth reduction, clipping, or codec artifacts. In practical deployments, access to clean speech target signals or labeled training data is unavailable, motivating the adaptation of a supervised SE to the test distribution.

Existing adaptation strategies include supervised fine-tuning, unsupervised domain adaptation (UDA), test-time training (TTT), and test-time adaptation (TTA) \cite{Tent}. Fine-tuning and UDA require access to the labeled source data \cite{UDA_OT, UDA_SSL}, while TTT constrains the original supervised training pipeline with auxiliary loss functions \cite{TTT, TTT_SSL}. In contrast, TTA methods update the pretrained model during inference using unsupervised loss functions, without modifying the original supervised training procedure \cite{TTA_PSE, TTA_Sub, LaDen,raichle2026test}. A commonly used variant, referred to as single‑instance or -utterance TTA, adapts the model independently for each test sample and reinitializes the model to its pretrained parameters after each update, thereby mitigating catastrophic forgetting \cite{TTA_ASR}.

Domain adaptation has been extensively studied for classification tasks, where many methods exploit statistical properties of model outputs. Those include techniques such as entropy minimization \cite{Tent, TTA_ASR, zhang2022memo} and feature alignment \cite{FAEastwoodMWS22, FAKojimaMI22}. However, such approaches do not readily translate to SE, which is typically formulated as a regression problem rather than a classification task.

Early approaches to domain adaptation for SE drew inspiration from the TTT paradigm and generally combined labeled source data with unlabeled target data through adversarial learning \cite{adversarialLiao0LW19, MichelsantiT17} or reconstruction‑based consistency objectives \cite{UDA_OT, UDA_SSL}. These methods require access to the source dataset during adaptation, which limits their practicality in real‑world deployment. 
The TTT-based method \cite{TTT} incorporates an auxiliary self-supervised loss during supervised training, which is then used only for adaptation on the unlabeled target data. Although effective, this method constrains the original training pipeline and cannot be applied to arbitrary pretrained SE models. This is not the case for the method proposed in \cite{UDA_SSL}, which explores the use of self-supervised models for speech representation learning. This approach performs adaptation by identifying samples in the source dataset whose latent representations are closest to those of the target unlabeled samples. However, this approach requires access to the source dataset during adaptation.

More recent research has focused on TTA methods that operate using exclusively unlabeled target data. RemixIT \cite{tzinis2022remixit} and LaDen~\cite{LaDen} both rely on a teacher-student framework where a student model is trained on the target data using pseudo-clean speech labels obtained with a teacher model. The method proposed in \cite{raichle2026test} introduces a single-utterance TTA method for SE methods based on time-frequency mask prediction. The approach fosters predicted masks to approach either $0$ or $1$ by minimizing their entropy, which encourages the model to produce more confident and discriminative masks for unseen noise conditions.

\begin{figure*}[h]
    \centering
    \includegraphics[width=0.85\linewidth]{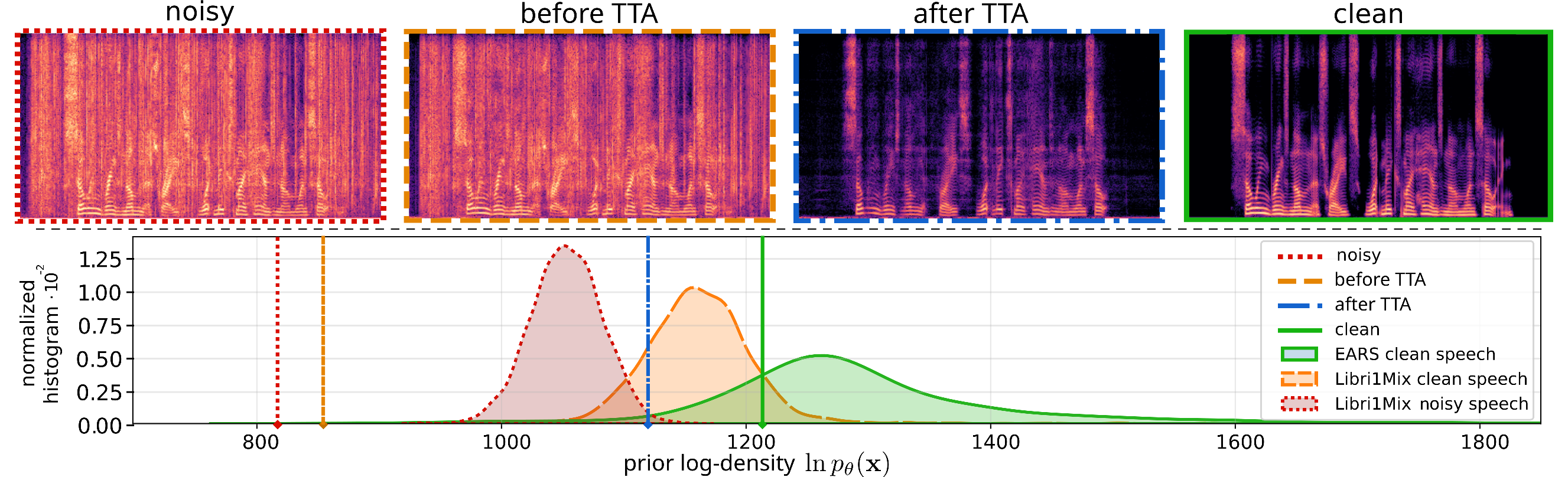}
    \caption{
    Visualization of the clean‑speech prior log‑density and its role in TTA. 
    \textbf{Top}: Spectrograms of a noisy input signal, the output of the pretrained SE model before and after adaptation, and the clean reference speech. \textbf{Bottom}: Normalized histogram of log‑density values for the pretrained prior evaluated on clean and noisy speech datasets. Vertical lines indicate the log-density values for the examples shown on top. The proposed TTA method pushes the enhanced output toward higher prior log-density values and closer to the clean reference.}
    \label{fig1}
    \vspace{-10pt}
\end{figure*}

In this paper, we propose a novel single-utterance TTA method for SE that leverages an autoregressive prior trained on clean speech latent representations extracted from a pretrained neural audio codec (NAC).  The core idea is to use the prior as a measure of how likely a given enhanced representation resembles clean speech, and to adapt the pretrained SE model by minimizing the Kullback–Leibler (KL) divergence between the distribution of the enhanced speech and the prior. As illustrated in Figure \ref{fig1}, a supervised SE model before adaptation can generalize poorly on unseen noise types, providing an output whose log-density under the prior is substantially lower than that of clean speech. Through TTA with the KL divergence, the enhanced output is progressively pushed toward regions of higher prior log-density, effectively encouraging it to align with the structure learned from clean data.
Experimental results on multiple real and synthetic noisy speech benchmarks show that the proposed method consistently improves speech quality, especially when the pretrained SE model operates under mismatched noise conditions.

\section{Method}
\vspace{-5pt}

\subsection{Supervised speech enhancement}

Supervised SE assumes access to a labeled dataset of parallel noisy‑clean speech recordings $\mathcal{D}_{\mathbf{x}\mathbf{y}} = \{ (\mathbf{x}_i, \mathbf{y}_i) \}_{i=1}^{N}$, where $\mathbf{x} \in \mathcal{A}$ and $\mathbf{y} \in \mathcal{A}$ denote the clean and noisy speech signals respectively, defined in some representation domain $\mathcal{A}$. SE can be cast as probabilistic inference using a conditional probability density function (PDF) ${q_\phi(\mathbf{x} \mid \mathbf{y})}$ with parameters $\phi$. The predicted clean speech is then obtained by computing the mean or $\arg\max$ of this inference model, or by sampling, given an estimate of the model parameters $\phi$. In a supervised training stage, those are learned by minimizing the negative log‑likelihood averaged over the labeled source dataset:
\begin{equation}
    \mathcal{L}_{\text{sup}}(\mathcal{D}_{\mathbf{x}\mathbf{y}}; \phi) = \frac{1}{N} \sum\nolimits_{(\mathbf{x}, \mathbf{y}) \in \mathcal{D}_{\mathbf{x}\mathbf{y}}} - \ln q_\phi(\mathbf{x} \mid \mathbf{y}).
    \label{eq:supervised_loss}
\end{equation}

The present work builds upon a fully supervised SE system operating in the latent space of a pretrained NAC \cite{SE_model}. Both clean and noisy waveforms are first encoded into NAC's latent representations before quantization, denoted by $\mathbf{x} = \{ \mathbf{x}_t \in \mathbb{R}^L \}_{t=1}^T$ and $\mathbf{y} = \{ \mathbf{y}_t  \in \mathbb{R}^L \}_{t=1}^T$, which serve as output targets and inputs for the SE model. Here, $L$ denotes the NAC latent space dimension and $T$ the number of time frames. The inference model is then defined as
\begin{equation}
q_\phi(\mathbf{x} \mid \mathbf{y}) =
\prod_{t=1}^T \mathcal{N}\big( \mathbf{x}_t ; f_{\phi,t} (\mathbf{y}), \mathbf{I} \big),
\label{eq:inference_model}
\end{equation}
where $f_\phi$ is a non‑autoregressive Conformer network \cite{SE_model, conformer}, and $\mathcal{N}(\mathbf{x}; \boldsymbol{\mu}, \boldsymbol{\Sigma})$ denotes the PDF of the multivariate Gaussian distribution, whose logarithm is given, up to an additive constant, by
\begin{equation}
\ln \mathcal{N}(\mathbf{x}; \boldsymbol{\mu}, \boldsymbol{\Sigma})
\overset{c}{=}
-\frac{1}{2}
\left[ \ln \mathrm{det}(\boldsymbol{\Sigma}) +
(\mathbf{x}-\boldsymbol{\mu})^\top \boldsymbol{\Sigma}^{-1} (\mathbf{x}-\boldsymbol{\mu})\right].
\label{eq:log_pdf_gaussian}
\end{equation}
Under the Gaussian model \eqref{eq:inference_model}, $\mathcal{L}_{\text{sup}}$ in \eqref{eq:supervised_loss} reduces to the mean{‑} squared error (MSE) loss function. It is clear from \eqref{eq:supervised_loss} that the parameters $\phi$ minimizing $\mathcal{L}_{\text{sup}}$ depend on the dataset $\mathcal{D}_{\mathbf{x}\mathbf{y}}$. While this supervised setting has been shown to be very effective, it assumes that the labeled training data capture the variability of acoustic environments encountered at test time. In practice, this assumption does not always hold, which can lead to poor generalization under strongly mismatched training and testing conditions \cite{pandey2020cross, bie2022unsupervised, richter2023speech, gonzalez2023assessing}.

\subsection{Unsupervised test-time adaptation}

The goal of this work is to improve enhancement performance on previously unseen noisy recordings, without access to additional parallel clean–noisy training data. To this end, we introduce an autoregressive prior trained on a clean speech corpus. TTA is then performed by optimizing the inference model parameters $\phi$ on a single noisy utterance using only the KL divergence between the inference model and the pretrained clean speech prior.\\\vspace{-5pt}

\myparagraph{Autoregressive prior} The clean speech prior $p_{\theta }(\mathbf{x})$ is defined as an autoregressive Gaussian model in the NAC latent space:
\begin{equation}
\label{Prior_def}
p_{\theta}(\mathbf{x})
 =
\prod_{t=1}^T 
p_{\theta}(\mathbf{x}_t \mid \mathbf{x}_{<t})
=\prod_{t=1}^T
\mathcal{N}\big(
\mathbf{x}_t ;
\boldsymbol{\mu}_{\eta ,t}(\mathbf{x}_{<t}),
\boldsymbol{\Sigma}_{\theta ,t}(\mathbf{x}_{<t})
\big),
\end{equation}
where $\mathbf{x}_{<t} = \{\mathbf{x}_s\}_{s=1}^t$, with the convention that $\mathbf{x}_{<1} = \emptyset$, and the covariance matrix is parameterized as follows:
\begin{equation}
    \boldsymbol{\Sigma}_{\theta ,t}( \mathbf{x}_{<t}) =
 \mathbf{W} \, \mathrm{diag}\{\mathbf{v}_{\eta ,t}(\mathbf{x}_{<t})\} \, \mathbf{W}^\top  \in \mathbb{R}^{L \times L},
 \label{eq:covariance_prior}
\end{equation}
with $\theta = \{\mathbf{W} \in \mathbb{R}^{L \times L}, \eta\}$. In this model, $\mathbf{W}$ is a global orthogonal matrix shared across time steps and speech signals, while the mean and variance vectors $\boldsymbol{\mu}_{\eta,t}( \mathbf{x}_{<t}), \mathbf{v}_{\eta,t}(\mathbf{x}_{<t})\in \mathbb{R}^L$ are obtained using a neural network at each time step $t$. 
The parametrization in \eqref{eq:covariance_prior} can be interpreted as learning a global eigenbasis for the NAC latent space, represented by the orthogonal matrix $\mathbf{W}$, while allowing the corresponding eigenvalues (the variances along each eigenvector's direction, represented by $\mathbf{v}_{\eta,t}(\mathbf{x}_{<t})$) to vary over time steps and conditioning speech signals. This model, inspired by \cite{muller2024post}, reflects the assumption that cross‑dimensional correlations are intrinsic to the latent space induced by the NAC encoder. It allows us to consider a full covariance matrix in potentially high dimension $L$, while remaining computationally efficient. Empirically, this full covariance model was found to lead to much higher log-density values than a diagonal covariance model when fitted on clean speech data. Given a clean speech corpus $\mathcal{D}_{\mathbf{x}} = \{ \mathbf{x}_i \in \mathbb{R}^{L \times T} \}_{i=1}^{M}$, the prior parameters $\theta$ are learned by minimizing the average negative log-likelihood using teacher forcing \cite{lamb2016professor}:
\begin{equation}
    \mathcal{L}_{\text{prior}} (\mathcal{D}_{\mathbf{x}}; \theta) =
    - \frac{1}{M} \sum_{\mathbf{x} \in \mathcal{D}_{\mathbf{x}}} \sum_{t=1}^T \ln  \mathcal{N}\big(
        \mathbf{x}_t ;
        \boldsymbol{\mu}_{\eta ,t}( \mathbf{x}_{<t}),
        \boldsymbol{\Sigma}_{\theta ,t}( \mathbf{x}_{<t})
\big),
\end{equation}
which can be easily computed using \eqref{eq:log_pdf_gaussian}.\\ 

\myparagraph{TTA loss function} At test time, given a noisy speech signal $\mathbf{y}$, the pretrained and frozen autoregressive prior $p_{\theta}(\mathbf{x})$ defined in \eqref{Prior_def}, and the pretrained SE model $q_\phi(\mathbf{x} \mid \mathbf{y})$ defined in \eqref{eq:inference_model}, adaptation is performed by optimizing with respect to $\phi$ the following TTA loss function based on the KL divergence:
\begin{align}
&\mathcal{L}_{\text{TTA}}(\mathbf{y}; \phi) = D_{\text{KL}}\big( q_\phi(\mathbf{x} \mid \mathbf{y}) \,\|\, p_{\theta }(\mathbf{x}) \big) \nonumber\\
&= \sum_{t=1}^T\mathbb{E}_{q_\phi(\mathbf{x}_{<t}\mid \mathbf{y})}   
            \bigl[ 
            D_{\text{KL}}\left(q_\phi(\mathbf{x}_t \mid  \mathbf{y}) \parallel p_{\theta}(\mathbf{x}_t \mid \mathbf{x}_{<t}) \right)
            \bigr] \label{TTA_loss_tmp} \\
& \overset{c}{\approx} -\frac{1}{2}\sum_{t=1}^T 
    \left\| 
    \mathrm{diag}\{\mathbf{v}_{\eta ,t}(\tilde{\mathbf{x}}_{<t})\}^{-\frac{1}{2}} 
     \mathbf{W}^\top   
     \bigr( 
        \boldsymbol{\mu}_{\eta ,t}( \tilde{\mathbf{x}}_{<t})
        - f_{\phi,t}(\mathbf{y})
     \bigl) 
     \right\|_2^2,
    \label{eq:TTA_loss}
\end{align}
where $D_{\text{KL}}(q \parallel p) = \mathbb{E}_q[\ln q - \ln p]$ and the approximation in \eqref{eq:TTA_loss} originates from the fact that we replace the intractable expectation in \eqref{TTA_loss_tmp} by the use of $\tilde{\mathbf{x}}_{<t} = \mathbb{E}_{q_\phi(\mathbf{x}_{<t}\mid \mathbf{y})} [\mathbf{x}_{<t}]= f_{\phi,<t}(\mathbf{y})$. In practice, the parameters $\phi$ are treated as fixed when computing $\tilde{\mathbf{x}}_{<t}$, such that gradients of $\mathcal{L}_{\text{TTA}}$ with respect to $\phi$ are not backpropagated through the computation of $\tilde{\mathbf{x}}_{<t}$.\\

\myparagraph{Adaptation algorithm} Adaptation is performed independently for each test sample, following a single‑utterance TTA setting. The noisy speech waveform is first normalized so that its amplitude is between $-1$ and $1$, after which a contiguous one‑second segment is randomly selected and encoded using the NAC encoder to obtain the latent representation $\mathbf{y} \in \mathbb{R}^{L \times T}$. 
This sequence is passed through the pretrained SE model to produce the enhanced sequence $\tilde{\mathbf{x}} = \{f_{\phi,t}(\mathbf{y}) \in \mathbb{R}^L\}_{t=1}^T$. 
The clean speech prior is then evaluated on $\tilde{\mathbf{x}}$, giving the mean and variance parameters $\{\boldsymbol{\mu}_{\eta,t}(\tilde{\mathbf{x}}_{<t}), \mathbf{v}_{\eta,t}(\tilde{\mathbf{x}}_{<t})  \in \mathbb{R}^L\}_{t=1}^T$. 
These quantities, as well as the orthogonal matrix $\mathbf{W}$, are used to compute the TTA loss function in \eqref{eq:TTA_loss}, which is optimized with respect to the inference model parameters $\phi$ using a fixed number of gradient descent steps $K$ and a small learning rate. The adapted model $f_\phi$ is then used to produce the latent representation of the complete enhanced speech signal, which is typically longer than the one-second segment used for adaptation. Therefore, the proposed TTA technique can be viewed as a calibration of the SE model to new acoustic noise characteristics, requiring only 1 second of representative unlabeled noisy speech. The clean speech waveform is eventually reconstructed using the NAC decoder. 

Note that the model parameters $\phi$ are reinitialized to their supervised pretrained values before processing another noisy speech signal. Indeed, if we keep optimizing the same parameters $\phi$ as new noisy speech signals arrive, the inference model will eventually collapse to the prior and become independent of the input noisy speech. To mitigate this problem, the parameters $\phi$ are reinitialized, and the TTA loss is optimized with a small number of steps and a small learning rate. This strategy is expected to encourage the inference model to output cleaner speech signals without becoming independent of the noisy input.

\section{Experiments}

\subsection{Experimental Setup}

\myparagraph{Datasets} The supervised SE model is trained on the Libri1Mix dataset \cite{librimix}, following the experimental setup described in \cite{SE_model}. The clean speech prior is trained independently using the EARS dataset \cite{EARS}, which provides anechoic speech recordings and is disjoint from the SE training data.

To evaluate the proposed test‑time adaptation method under realistic and mismatched conditions, we primarily conduct experiments on the DNS Challenge V5 dev‑test set, track 2 \cite{DNSv5}. This dataset contains 600 real noisy speech recordings without parallel clean references and reflects a wide range of acoustic environments and noise conditions. To further assess generalization, we validate our approach on additional datasets with varying degrees of mismatch. These include the TIMIT‑DEMAND dataset \cite{leglaive2018variance}, which introduces both unseen speakers and unseen noise environments; the EARS‑WHAM dataset \cite{EARS}, which combines clean speech drawn from the same distribution as the prior training data with noise conditions seen during SE training; and the Libri1Mix test set, which represents a matched condition for the supervised SE model. This selection allows us to disentangle the effects of speaker mismatch, noise mismatch, and prior–SE alignment.\\\vspace{-5pt}

\myparagraph{Evaluation metrics} Since we are working in the latent space of a NAC trained with adversarial learning and without an explicit phase-aware reconstruction loss function, NAC outputs are not aligned sample-wise with the clean speech reference \cite{kumar2020nu, langman2024spectral}. Therefore, and as commonly done in the literature \cite{wang2024selm}, we primarily rely on the non‑intrusive DNSMOS P.835 metrics \cite{reddy2022dnsmos}, which include SIG (speech quality), BAK (background noise suppression), and OVRL (overall quality), each ranging from 1 to 5.
In addition, we use the word error rate (WER) as an intrusive intelligibility metric whenever reference transcripts are available. WER is computed using a pretrained wav2vec 2.0 ASR model\footnote{https://huggingface.co/facebook/wav2vec2-base-960h}, with the provided transcripts serving as ground truth.\\\vspace{-5pt}

\myparagraph{TTA setup} All experiments use the Descript Audio Codec (DAC) \cite{DAC} as the NAC architecture underlying the SE and prior models. TTA is performed using standard gradient descent with momentum $0.9$. The learning rate is fixed to $2 \times 10^{-5}$, and the maximum number of adaptation steps is empirically set to $K = 20$.
At every two adaptation steps, the enhanced speech is reconstructed using the adapted model parameters and evaluated using the selected metrics. This allows tracking the evolution of quality and intelligibility throughout the adaptation process. \\\vspace{-5pt}

 \begin{figure}
    \centering
    \includegraphics[width=0.8\linewidth]{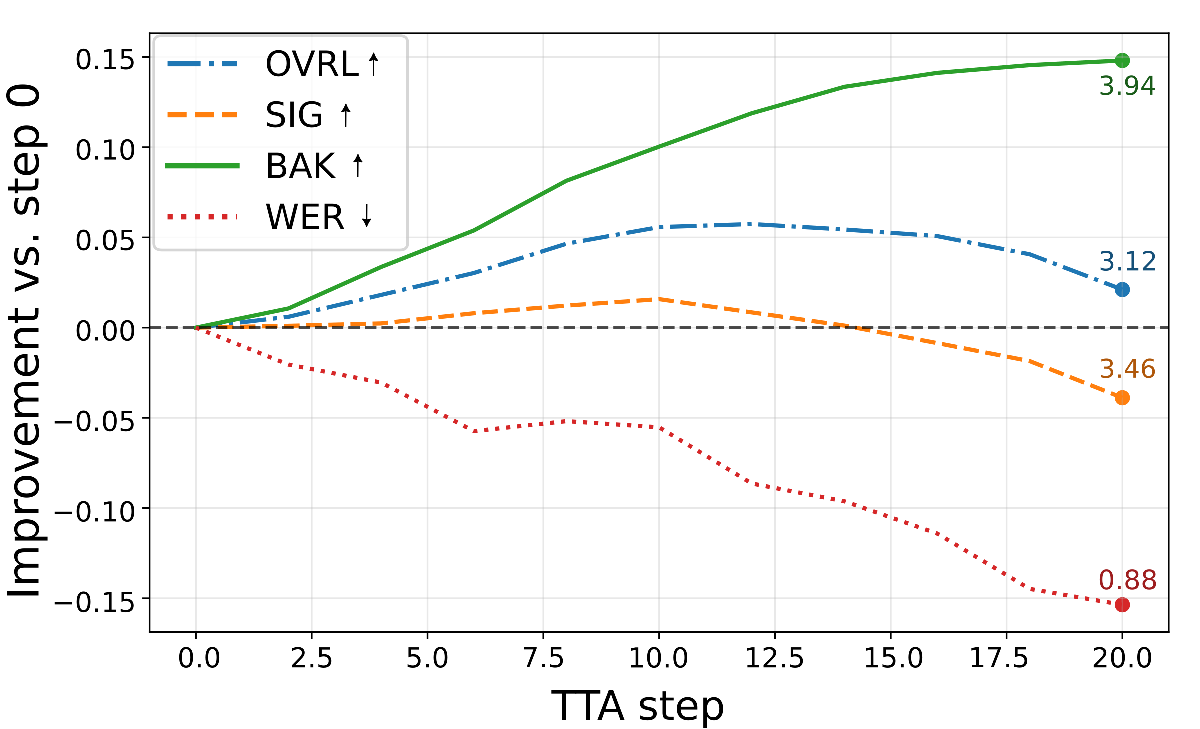}
    \caption{Relative improvement of the metrics along TTA iterations, computed on the DNS Challenge V5 dev‑test set.}
    \label{fig2}
\end{figure}

 \begin{figure}
    \centering
    \includegraphics[width=\linewidth]{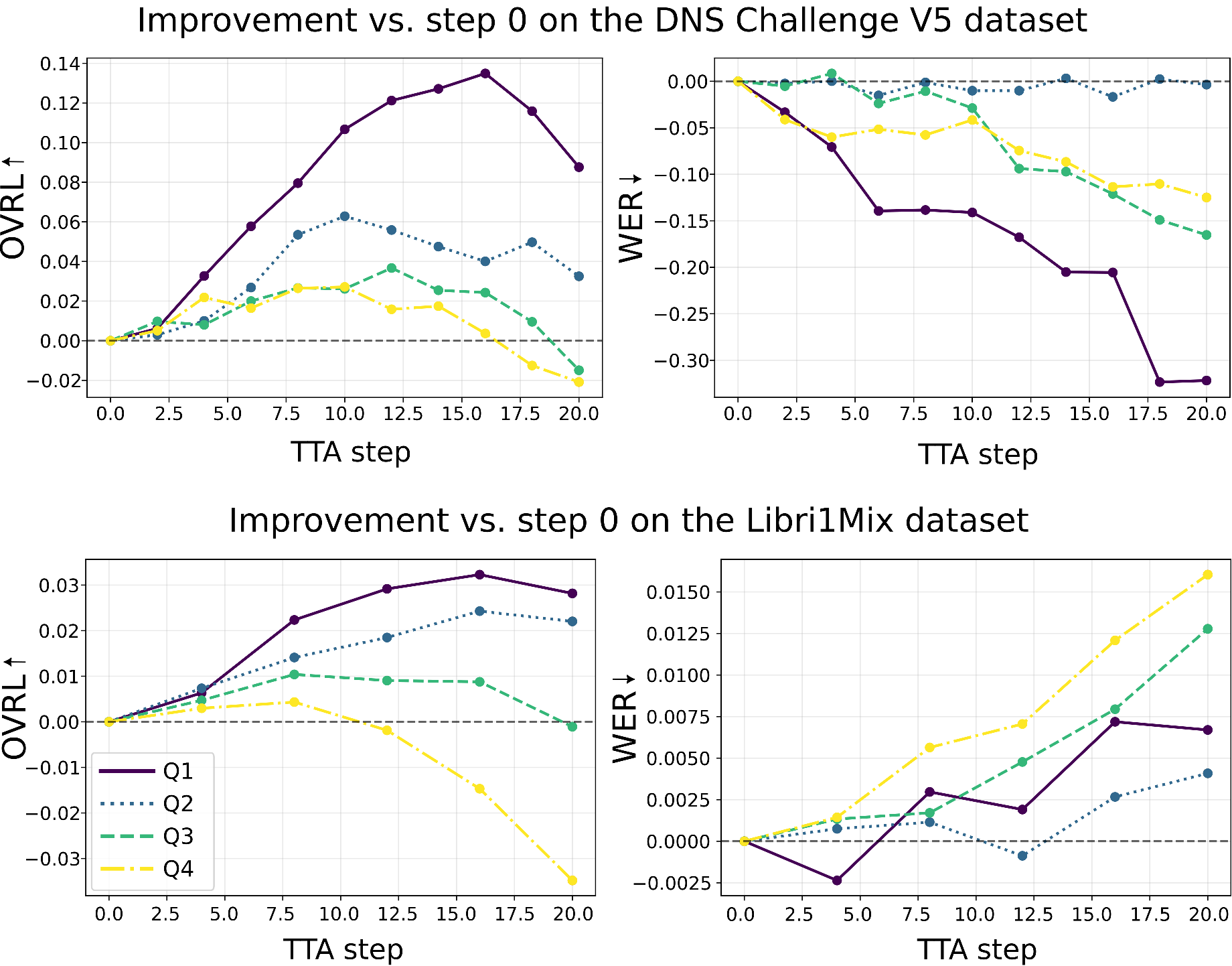}
    \caption{OVRL and WER relative improvements over TTA steps and per initial prior log-density quartile, for the DNS Challenge V5 and Libri1Mix datasets.}
    \label{fig3}
\vspace{-15pt}
    
\end{figure}

\subsection{Results}

\myparagraph{Prior model validation} The objective of the clean speech prior is to discriminate between clean and noisy speech by assigning higher log‑density values to clean latent speech representations. To validate this behavior, we evaluate the prior log‑density on several datasets and analyze its distribution.
Figure \ref{fig1} shows histograms of the log-density values computed on the EARS clean test set, as well as on the clean and noisy subsets of the Libri1Mix test set. The results indicate a clear separation between clean and noisy speech, with clean speech consistently assigned higher values. We also visualize the prior log-density values and the spectrograms of a noisy speech signal, the enhanced signal before and after TTA, and the clean reference. We chose an example for which the pretrained SE model completely fails to suppress background noise, because of a mismatch with the training noise types. This qualitative analysis confirms that the prior assigns a higher log-density to cleaner speech representations. These observations support the use of the prior as a form of weak supervision during TTA, where the objective is to push the enhanced speech representation toward regions of higher prior log-density in the latent space.\\
\vspace{-7pt}

\myparagraph{TTA results} We first analyze the behavior of the proposed TTA method on the DNS Challenge V5 dev‑test set. Figure \ref{fig2} reports the relative improvement of DNSMOS metrics and WER as a function of the adaptation step, compared to the model output at step $k = 0$, corresponding to the pretrained SE model without adaptation. On average, speech quality improves during the early adaptation steps, before eventually degrading as the number of steps increases. This behavior is expected, as the TTA objective relies exclusively on the unsupervised KL divergence, which may eventually cause the inference model to collapse to the prior and thus ignore the noisy speech input. These results highlight the necessity of early stopping mechanisms to prevent over‑adaptation.

Informal listening tests and quantitative analysis reveal that the effectiveness of TTA strongly depends on the initial quality of the enhanced speech. Specifically, TTA tends to yield larger improvements when the pretrained SE model leaves residual noise in the enhanced output, while it can degrade performance when the initial enhancement quality is already high. To validate this observation, we compute the prior log-density of the enhanced speech at step $k=0$ and partition the dataset into four equally sized quartiles, $Q_1$ to $Q_4$, sorted by increasing log-density values. The underlying intuition is that the prior log-density provides a proxy for enhancement quality, with lower values corresponding to noisier outputs. Figure~\ref{fig3} reports the evolution of OVRL and WER improvements over TTA iterations for each quartile on both the DNS V5 and Libri1Mix datasets. The results show that lower‑log-density quartiles benefit more significantly from adaptation and typically require more adaptation steps to reach their optimal performance. In contrast, higher‑log-density quartiles exhibit limited or negative gains, confirming the sensitivity of TTA to the initial operating point. Note that the WER increase on Libri1Mix is negligible, reaching at most about 0.015 for $Q_4$.

The optimal step can be computed as $k^* = \arg\max_k \mathrm{OVRL}(k)$ using the non-intrusive DNSMOS metric, but this approach is computationally expensive. Motivated by the quartile analysis, we investigate training a simple logistic regression model that gives a prediction $\tilde{k}$ of $k^*$ based solely on the prior log-density of the enhanced speech before TTA.  The regression model is trained using the DNS Challenge results and then applied to other datasets. 

The results reported in Table \ref{tab:tta_results} show improvements across all datasets for $k=k^*$, and also for $k = \tilde{k}$, although to a lesser extent. Performance gains are more important for the DNS Challenge and TIMIT-DEMAND datasets, especially in terms of BAK score. This is because those datasets are mismatched to the one used for supervised pretraining of the SE model in terms of noise type. We can indeed see that the BAK and OVRL scores obtained by the pretrained SE model before TTA are lower on the DNS Challenge and TIMIT-DEMAND datasets than on EARS-WHAM and Libri1Mix, even though the noisy input scores are higher. Overall, this quantitative analysis confirms that the proposed TTA method is particularly effective at reducing residual background noise when the pretrained supervised SE model fails on unseen noise types, as also illustrated qualitatively in Figure~\ref{fig1} and in the online audio examples.$^1$






\begin{table}[t]
\centering
\caption{Test-time adaptation results across datasets. ``SE w/o TTA $(k=0)$'' denotes to the pretrained SE model before TTA, while ``SE w/ TTA $(k=\tilde{k} / k^*)$'' denotes the adapted model at different steps.\vspace{3pt}}
\label{tab:tta_results}
\setlength{\tabcolsep}{5pt}
\renewcommand{\arraystretch}{1} 
\fontsize{8.5pt}{10pt}\selectfont 
\scalebox{0.9}
{
\begin{tabular}{l l r r r}
\toprule
& \textbf{} & \textbf{OVRL} & \textbf{SIG} & \textbf{BAK} \\
\midrule

\multirow{4}{*}{\textbf{DNS Challenge}}
& Noisy input & $2.22$ & $3.19$ & $2.38$ \\
& SE w/o TTA $(k=0)$ & $3.09$ & $3.50$ & $3.80$ \\
& SE w/ TTA $(k=\tilde{k})$ & $+0.05$ & $-0.02$ & $+0.15$ \\
& SE w/ TTA $(k=k^*)$ & $+0.18$ & $+0.08$ & $+0.23$ \\
\midrule

\multirow{4}{*}{\textbf{TIMIT-DEMAND}}
& Noisy input & $2.24$ & $2.89$ & $2.69$ \\
& SE w/o TTA $(k=0)$ & $3.00$ & $3.51$ & $3.62$ \\
& SE w/ TTA $(k=\tilde{k})$ & $+0.07$ & $+0.02$ & $+0.13$ \\
& SE w/ TTA $(k=k^*)$ & $+0.15$ & $+0.06$ & $+0.22 $\\
\midrule

\multirow{4}{*}{\textbf{EARS-WHAM}}
& Noisy input & $2.09$ & $2.72$ & 2.39 \\
& SE w/o TTA $(k=0)$ & $3.24$ & $3.50$ & 3.99 \\
& SE w/ TTA $(k=\tilde{k})$ &$ 0.00$ & $-0.01$ & $+0.02$ \\
& SE w/ TTA $(k=k^*)$ & $+0.09$ & $+0.08$ & $+0.09$ \\
\midrule

\multirow{4}{*}{\textbf{Libri1Mix}}
& Noisy input & $1.75$ & $2.46$ & $1.81$ \\
& SE w/o TTA $(k=0)$ & $3.28$ & $3.57$ & $4.03$ \\
& SE w/ TTA $(k=\tilde{k})$ & $+0.03$ & $+0.02$ & $+0.03$ \\
& SE w/ TTA $(k=k^*)$ & $+0.08$ & $+0.05$ & $+0.07$ \\
\bottomrule

\end{tabular}
}
\vspace{-12pt}
\end{table}

\vspace{-7pt}
\section{Conclusion}
\vspace{-7pt}

We presented a novel single-utterance TTA framework for SE that leverages an autoregressive clean speech prior defined in the latent space of a NAC. By optimizing a pretrained SE model through a KL divergence objective, the proposed method enables unsupervised adaptation to unseen acoustic conditions without modifying the original training pipeline or requiring access to source data.
Experimental results demonstrated that the clean speech prior effectively discriminates between noisy and clean speech and serves as a meaningful adaptation signal. 
Future work will explore more expressive prior models and extensions to non-additive distortions.

\cleardoublepage

\bibliographystyle{IEEEtran}
\begingroup
\fontsize{9pt}{10.5pt}\selectfont 
\bibliography{refs}

@string{TASLP_new = "IEEE/ACM Trans. Audio Speech Lang. Process."}

@string{JSTSP = "IEEE J. Sel. Top. Signal Process."}

@string{OJSP = "IEEE Open J. Signal Process."}

@string{MLSP = "IEEE Int. Workshop Mach. Learn. Signal Process. (MLSP)"}

@string{ICASSP = "IEEE Int. Conf. Acoust. Speech Signal Process. (ICASSP)"}

@string{WASPAA = "IEEE Workshop Appl. Signal Process. Audio Acoust. (WASPAA)"}

@string{EUSIPCO = "Eur. Signal Process. Conf. (EUSIPCO)"}

@string{INTERSPEECH = "Interspeech"}

@string{IJCAI = "Int. Jt. Conf. Artif. Intell. (IJCAI)"}

@string{NeurIPS = "Adv. Neural Inf. Process. Syst. (NeurIPS)"}

@string{ICML = "Int. Conf. Mach. Learn. (ICML)"}

@string{ICLR = "Int. Conf. Learn. Represent. (ICLR)"}

@inproceedings{wang2024selm,
  title={{SELM}: Speech enhancement using discrete tokens and language models},
  author={Wang, Ziqian and Zhu, Xinfa and Zhang, Zihan and Lv, YuanJun and Jiang, Ning and Zhao, Guoqing and Xie, Lei},
  booktitle=ICASSP,
  year={2024}
}

@article{langman2024spectral,
  title={Spectral codecs: Spectrogram-based audio codecs for high quality speech synthesis},
  author={Langman, Ryan and Jukic, Ante and Dhawan, Kunal and Koluguri, Nithin Rao and Ginsburg, Boris},
  journal={arXiv preprint arXiv:2406.05298},
  year={2024}
}

@article{kumar2020nu,
  title={{NU-GAN}: High resolution neural upsampling with {GAN}},
  author={Kumar, Rithesh and Kumar, Kundan and Anand, Vicki and Bengio, Yoshua and Courville, Aaron},
  journal={arXiv preprint arXiv:2010.11362},
  year={2020}
}

@inproceedings{leglaive2018variance,
  title={A variance modeling framework based on variational autoencoders for speech enhancement},
  author={Leglaive, Simon and Girin, Laurent and Horaud, Radu},
  booktitle=MLSP,
  year={2018}
}

@article{gonzalez2023assessing,
  title={Assessing the generalization gap of learning-based speech enhancement systems in noisy and reverberant environments},
  author={Gonzalez, Philippe and Alstr{\o}m, Tommy Sonne and May, Tobias},
  journal=TASLP_new,
  volume={31},
  pages={3390--3403},
  year={2023},
}

@article{pandey2020cross,
  title={On cross-corpus generalization of deep learning based speech enhancement},
  author={Pandey, Ashutosh and Wang, DeLiang},
  journal=TASLP_new,
  volume={28},
  pages= {2489--2499},
  year={2020},
}

@article{richter2023speech,
  title={Speech enhancement and dereverberation with diffusion-based generative models},
  author={Richter, Julius and Welker, Simon and Lemercier, Jean-Marie and Lay, Bunlong and Gerkmann, Timo},
  journal=TASLP_new,
  volume={31},
  pages={2351--2364},
  year={2023}
}

@article{bie2022unsupervised,
  title={Unsupervised speech enhancement using dynamical variational autoencoders},
  author={Bie, Xiaoyu and Leglaive, Simon and Alameda-Pineda, Xavier and Girin, Laurent},
  journal=TASLP_new,
  volume={30},
   pages={2993--3007},
  year={2022},
}

@article{lamb2016professor,
  title={Professor forcing: A new algorithm for training recurrent networks},
  author={Lamb, Alex M and Goyal, Anirudh and Zhang, Ying and Zhang, Saizheng and Courville, Aaron C and Bengio, Yoshua},
  journal=NeurIPS,
  volume={29},
  year={2016}
}

@inproceedings{conformer,
title={Conformer: Convolution-augmented transformer for speech recognition},
author={Gulati, A. and Qin, J. and Chiu, C. and Parmar, N. and Zhang, Y. and Yu, J. and Han, W. and Wang, S. and Zhang, Z. and Wu, Y. and others},
booktitle=INTERSPEECH,
year={2020}
}

@article{tzinis2022remixit,
  title={Remixit: Continual self-training of speech enhancement models via bootstrapped remixing},
  author={Tzinis, Efthymios and Adi, Yossi and Ithapu, Vamsi K and Xu, Buye and Smaragdis, Paris and Kumar, Anurag},
  journal=JSTSP,
  volume={16},
  year={2022},
}

@inproceedings{TTT,
  title     = {{Test-Time Training for Speech Enhancement}},
  author    = {Avishkar Behera and
               Riya Ann Easow and
               Venkatesh Parvathala and
               K. Sri Rama Murty},
  year      = {2025},
  booktitle = INTERSPEECH,
}

@INPROCEEDINGS{UDA_SSL,
  author={Lee, Ching-Hua and Yang, Chouchang and Srinivasa, Rakshith Sharma and Malur Saidutta and others},
  booktitle=ICASSP, 
  title={Leveraging Self-Supervised Speech Representations for Domain Adaptation in Speech Enhancement}, 
  year={2024},
  }

@inproceedings{UDA_OT,
  author       = {Hsin{-}Yi Lin and
                  Huan{-}Hsin Tseng and
                  Xugang Lu and
                  Yu Tsao},
  title        = {Unsupervised Noise Adaptive Speech Enhancement by Discriminator-Constrained Optimal Transport},
  booktitle    = NeurIPS,
  year         = {2021}
}

@article{LaDen,
  title={Test-Time Adaptation for Speech Enhancement Via Domain Invariant Embedding Transformation},
  author={Raichle, Tobias and Edinger, Niels and Yang, Bin},
  journal=OJSP,
  year={2026},
  publisher={IEEE}
}

@inproceedings{raichle2026test,
  title={Test-Time Adaptation For Speech Enhancement Via Mask Polarization},
  author={Raichle, Tobias and Amini, Erfan and Yang, Bin},
  booktitle=ICASSP,
  year={2026}
}

@inproceedings{Tent,
  author       = {Dequan Wang and
                  Evan Shelhamer and
                  Shaoteng Liu and
                  Bruno A. Olshausen and
                  Trevor Darrell},
  title        = {Tent: Fully Test-Time Adaptation by Entropy Minimization},
  booktitle    = ICLR,
  year         = {2021}
}

@inproceedings{TTT_SSL,
  author       = {Yu Sun and
                  Xiaolong Wang and
                  Zhuang Liu and
                  John Miller and
                  Alexei A. Efros and
                  Moritz Hardt},
  title        = {Test-Time Training with Self-Supervision for Generalization under
                  Distribution Shifts},
  booktitle    = ICML,
  year         = {2020}
  }

@inproceedings{TTA_ASR,
  author       = {Guan{-}Ting Lin and
                  Shang{-}Wen Li and
                  Hung{-}yi Lee},
  title        = {Listen, Adapt, Better {WER:} Source-free Single-utterance Test-time
                  Adaptation for Automatic Speech Recognition},
  booktitle    = INTERSPEECH,
  year         = {2022}
}

@inproceedings{TTA_Sub,
  author       = {Kazuki Adachi and
                  Shin'ya Yamaguchi and
                  Atsutoshi Kumagai and
                  Tomoki Hamagami},
  title        = {Test-time Adaptation for Regression by Subspace Alignment},
  booktitle    =  ICLR,
  year         = {2025}
}

@inproceedings{TTA_PSE,
  author       = {Sunwoo Kim and
                  Minje Kim},
  title        = {Test-Time Adaptation Toward Personalized Speech Enhancement: Zero-Shot
                  Learning with Knowledge Distillation},
  booktitle    = WASPAA,
  year         = {2021}
}

@inproceedings{DNSv5,
  title={ICASSP 2023 Deep Noise Suppression Challenge},
  author={
 Dubey, Harishchandra and Aazami, Ashkan and Gopal, Vishak and Naderi, Babak and Braun, Sebastian and  Cutler, Ross and Gamper, Hannes and Golestaneh, Mehrsa and Aichner, Robert},
  booktitle=ICASSP,
  year={2023}
}

@article{zhang2022memo,
  title={Memo: Test time robustness via adaptation and augmentation},
  author={Zhang, Marvin and Levine, Sergey and Finn, Chelsea},
  journal=NeurIPS,
  year={2022}
}

@inproceedings{FAKojimaMI22,
  author       = {Takeshi Kojima and
                  Yutaka Matsuo and
                  Yusuke Iwasawa},
  title        = {Robustifying Vision Transformer without Retraining from Scratch by
                  Test-Time Class-Conditional Feature Alignment},
  booktitle    = IJCAI,
  year         = {2022}
}

@inproceedings{FAEastwoodMWS22,
  author       = {Cian Eastwood and
                  Ian Mason and
                  Christopher K. I. Williams and
                  Bernhard Sch{\"{o}}lkopf},
  title        = {Source-Free Adaptation to Measurement Shift via Bottom-Up Feature
                  Restoration},
  booktitle    = ICLR,
  year         = {2022}
}

@inproceedings{MichelsantiT17,
  author       = {Daniel Michelsanti and
                  Zheng{-}Hua Tan},
  title        = {Conditional Generative Adversarial Networks for Speech Enhancement
                  and Noise-Robust Speaker Verification},
  booktitle    = INTERSPEECH,
  year         = {2017}
}

@inproceedings{adversarialLiao0LW19,
  author       = {Chien{-}Feng Liao and
                  Yu Tsao and
                  Hung{-}yi Lee and
                  Hsin{-}Min Wang},
  title        = {Noise Adaptive Speech Enhancement Using Domain Adversarial Training},
  booktitle    = INTERSPEECH,
  year         = {2019}
}

@inproceedings{SE_model,
  author       = {Sofiene Kammoun and
                  Xavier Alameda{-}Pineda and
                  Simon Leglaive},
  title        = {Modeling strategies for speech enhancement in the latent space of
                  a neural audio codec},
  booktitle      = ICASSP,
  year         = {2026}
}

@inproceedings{muller2024post,
  title={Post-Training Latent Dimension Reduction in Neural Audio Coding},
  author={Muller, Thomas and Ragot, St{\'e}phane and Philippe, Pierrick and Scalart, Pascal},
  booktitle=EUSIPCO,
  year={2024},
}

@article{librimix,
author = {Cosentino, J. and others},
title = {{LibriMix}: An Open-Source Dataset for Generalizable Speech Separation},
journal = {arXiv:2005.11262},
year = {2020}
}

@inproceedings{EARS,
  author       = {Julius Richter and
                  Yi{-}Chiao Wu and
                  Steven Krenn and
                  Simon Welker and
                  Bunlong Lay and
                  Shinji Watanabe and
                  Alexander Richard and
                  Timo Gerkmann},
  title        = {{EARS:} An Anechoic Fullband Speech Dataset Benchmarked for Speech
                  Enhancement and Dereverberation},
  booktitle    = INTERSPEECH,
  year         = {2024}
}

@inproceedings{reddy2022dnsmos,
title={{DNSMOS P. 835}: A non-intrusive perceptual objective speech quality metric to evaluate noise suppressors},
author={Reddy, C. K. and Gopal, V. and Cutler, R.},
booktitle=ICASSP,
year={2022}
}

@inproceedings{DAC,
author = {Kumar, R. and Seetharaman, P. and Luebs, A. and Kumar, I. and Kumar, K.},
title= {High-Fidelity Audio Compression with Improved {RVQGAN}},
booktitle= NeurIPS,
year = {2023}
}
\endgroup
\end{document}